# Engineering LLM Powered Multi-agent Framework for Autonomous CloudOps

Kannan Parthasarathy*, Karthik Vaidhyanathan†, Rudra Dhar†, Venkat Krishnamachari*, Basil Muhammed*,
Adyansh Kakran†, Sreemaee Akshathala†, Shrikara Arun†, Sumant Dubey*, Mohan Veerubhotla*, Amey Karan†
*MontyCloud Inc
†Software Engineering Research Center, IIIT Hyderabad, India
Email: karthik.vaidhyanathan@iiit.ac.in, {kannan, venkat, basil, sumant, mohan}@montycloud.com,
{rudra.dhar, adyansh.kakran, sreemaee.akshathala, amey.karan}@research.iiit.ac.in, shrikara.a@students.iiit.ac.in

*Abstract*—Cloud Operations (CloudOps) is a rapidly growing field focused on the automated management and optimization of cloud infrastructure which is essential for organizations navigating increasingly complex cloud environments. MontyCloud Inc is one of the major companies in the CloudOps domain that leverages autonomous bots to manage cloud compliance, security, and continuous operations. To make their platform more accessible and effective to the customers, MontyCloud worked with us to leverage the use of GenAI.

Developing a GenAI-based solution for autonomous CloudOps for the existing MontyCloud system presented us with various challenges such as i) diverse data sources; ii) orchestration of multiple processes and iii) handling complex workflows to automate routine tasks. To this end, we developed MOYA, a multi-agent framework that leverages GenAI and balances autonomy with the necessary human control. This framework integrates various internal and external systems and is optimised for factors like task orchestration, security, and error mitigation while producing accurate, reliable and relevant insights by utilising Retrieval Augmented Generation (RAG). Evaluations of our multi-agent system with the help of practitioners as well as using automated checks demonstrate enhanced accuracy, responsiveness, and effectiveness over non-agentic approaches across complex workflows.

## I. INTRODUCTION

Cloud computing has fundamentally transformed the IT landscape, providing organisations of all sizes with instant access to scalable, flexible, and cost-effective digital infrastructure. Businesses increasingly use cloud platforms for storage, application hosting, analytics, and AI. Today, many organisations are adopting cloud solutions. According to a survey, over 94% of organisations operate cloud-based workloads, that is applications, processes, or tasks running on cloud infrastructure and utilising resources like compute, storage, and network [1],[2]. However, this widespread adoption introduces the shared responsibility model [3], where both cloud providers and customers have roles in managing the cloud environment. It places operational responsibilities on organisations, requiring a well-defined CloudOps practice to effectively manage their share of duties.

CloudOps, or Cloud Operations, refers to the practices, tools, and processes to manage, optimise, and secure applications and infrastructure in the cloud. Alonso et al. [1] defines it as a framework that extends *DevOps* practices to cloud management by adding components like resource discovery, self-healing, and real-time monitoring. By focusing on automation, monitoring, cost management, and compliance, CloudOps enables organisations to maintain efficient, resilient, and scalable cloud environments. However, the complex and dynamic nature of cloud services makes manual management time-intensive, challenging, and prone to errors.

MontyCloud's Autonomous CloudOps platform addresses these challenges by automating workflows to streamline operations and provide real-time visibility into inventory, security, and costs [4]. The platform tackles challenges such as navigating the complexity of hundreds of services, establishing secure and cost-effective cloud governance, ensuring a strong security posture, and adhering to evolving compliance standards.

The rise of AI, particularly Generative AI (GenAI) and Large Language Models (LLMs), is transforming businesses and streamlining operations. Hou et al. [2] outline the successful applications of LLMs in various software engineering tasks, highlighting their growing significance in the field. CloudOps can also benefit from this emerging technology. Hence, to enhance the accessibility of their platform, and to streamline cloud operations, MontyCloud worked with us to develop an Autonomous CloudOps solution that leverages GenAI in particular LLMs.

However, engineering GenAI systems for CloudOps is challenging due to the need for real-time processing of diverse data streams while ensuring reliability and quality. Generative models often lack context, leading to inaccuracies in complex domains like CloudOps. Retrieval-Augmented Generation (RAG) [3] addresses this by integrating information retrieval, enhancing relevance and accuracy. Initially, we developed a monolithic RAG system combining a retrieval engine and a generative model to improve MontyCloud's platform accessibility. As features expanded, the system faced challenges in handling large data volumes from diverse sources and integrating various in-house and external APIs, exposing the limitations of our monolithic architecture. Multiple LLM calls risk context overload and reduced accuracy as they lack domain-specific focus in processing [4]. Emerging multi-agent systems have the potential to address these challenges. However, existing agentic frameworks are unsuitable for our CloudOps needs. To this end, we developed a novel Multi-agent-based GenAI framework, MOYA (Meta Orchestrator Of Your Agents). It addresses the limitations of existing frameworks by balancing automation with necessary human supervision, ensuring efficiency, extensibility, and security in our CloudOps operations.

---

[1] https://resources.flexera.com/web/media/documents/rightscale-2019-state-of-the-cloud-report-from-flexera.pdf
[2] https://www.fortinet.com/content/dam/fortinet/assets/analyst-reports/report-2022-cloud-security.pdf
[3] https://aws.amazon.com/compliance/shared-responsibility-model/
[4] https://montycloud.com

In this work, we provide the details of the framework, explaining the different components and how they interact with each other. We explore the shift from a monolithic system to a multi-agent framework in CloudOps and empirically assess their performance. By utilising both standard metrics and human evaluations, we demonstrate that the proposed framework outperformed the traditional RAG-based approach that relied on a single LLM. Figure 1 shows an example scenario where multiple agents collaborate to resolve user queries instead of a single LLM performing the entire task.

The rest of the paper is structured as follows: Section II provides a background on CloudOps and Multi-Agent frameworks. Section III outlines the key challenges in CloudOps and limitations of our initial monolithic approach, Section IV details the architecture of MOYA Framework whereas Section V describes some key agents. Section VI presents the experimental setup and evaluation methodologies employed. In Section VII, we interpret the findings of our study and its relevance to the community. Section IX gives an overview of the relevant literature and current works and Section X provides conclusions and insights into future works.

## II. BACKGROUND

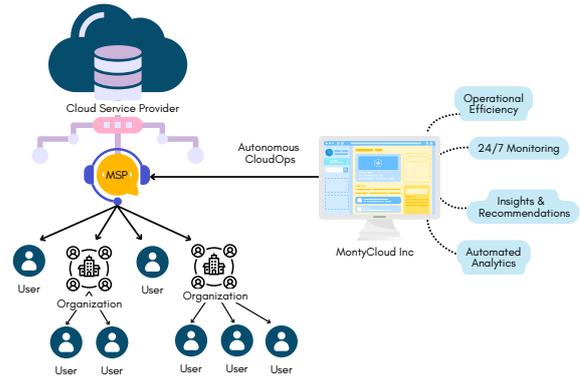

Fig. 2. Autonomous CloudOps

### A. Autonomous CloudOps of MontyCloud

Efficient CloudOps management is essential for organisations to fully capitalise on the cloud's agility, scalability, and potential for innovation. These operations are provided to the organisations by Managed Service Providers (MSPs). MSPs bring extensive expertise in CloudOps and maintain strong partnerships with cloud providers. They assist organisations at every stage of their CloudOps lifecycle, including planning, migration, operations, and optimisation.

However, managing CloudOps at scale presents unique challenges for MSPs who often oversee operations for numerous organisations, each with unique needs and intricate infrastructures. This volume demands constant monitoring, analysis, and optimisation, which can become unsustainable when relying on manual management. Such inefficiencies restrict the ability to deliver timely insights and implement optimisations effectively.

To address these challenges, MontyCloud developed a system that can support Autonomous CloudOps, as shown in Figure 2. This system provides services like real-time visibility, continuous monitoring, and data-driven automation to the MSPs for automating the management of CloudOps for various users and organisations. In doing so, MontyCloud enables MSPs to deliver reliable and optimised cloud services across different verticals like security, governance, compliance, and cost management at scale.

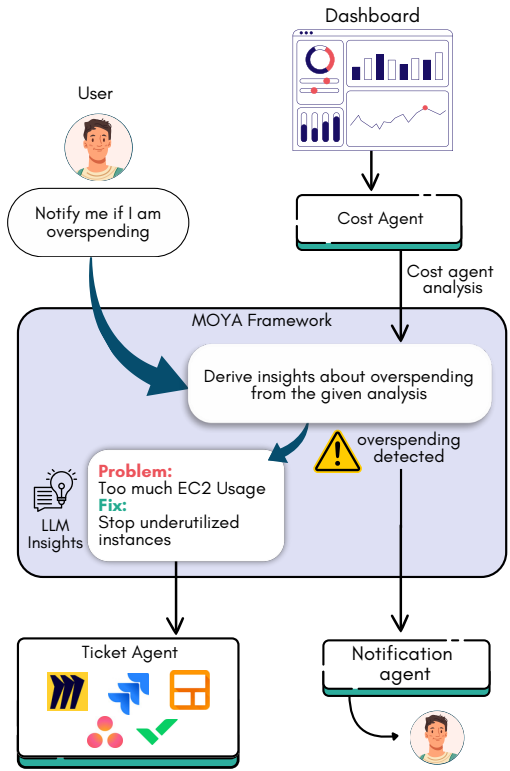

Fig. 1. Multiple agents collaborating to process user queries

*B. Multi-Agent Systems*

A Multi-Agent System (MAS) [5] is a framework that uses multiple intelligent agents to autonomously solve complex problems that may be difficult for a single agent or centralised system. Rooted in game theory, MAS leverages intelligent entities customised for specific tasks to improve scalability and adaptability in various applications. In software engineering, MAS frameworks enable decentralised decision-making and efficient collaboration among agents [6]. Integrating large language models (LLMs) into these frameworks has expanded their capabilities, with systems like AutoGen [7], CrewAI [5], LangGraph [6], and AutoGPT [7] offering diverse approaches and benefits. For instance, LangGraph uses a design-time graph structure, which demands individual customisations for different customers, restricting run-time flexibility. While frameworks like Autogen, AutoGPT, and CrewAI, allow for runtime decision-making, the level of abstraction limited our ability to customise interactions and integrate with MontyCloud's unique data sources and APIs. While custom tools could be built within these existing frameworks, creating a tailored multi-agent framework provided us with greater flexibility and control to meet our CloudOps needs and scalability goals. It also enabled seamless integration with MontyCloud's data and APIs, while allowing external developers to define and deploy their own agents.

## III. CHALLENGES

The application of AI in cloud operations is still relatively new and underdeveloped. Comprehensive studies by Martinez et al. [8] and Bosch et al. [9] identify key challenges in engineering AI-based systems. These challenges include data-related issues, difficulties in understanding qualitative attributes, the applicability of software engineering practices, and maintenance problems associated with building AI systems. The introduction of GenAI in CloudOps makes it more challenging to establish a system that incorporates Software Engineering best practices, especially for real-time decision-making, consistency, and adaptability.

Our initial design relied on RAG as the backbone to address some of these foundational challenges. Although a RAG-based monolithic system significantly reduced the need for manual analysis by facilitating real-time data access, issue identification, anomaly detection, automated chat-based responses, and accurate insights, it also brought its own set of challenges as the system evolved. The core challenges we recognised with our initial system are:

*A. CH1: Managing Distributed Data*

Cloud environments generate large amounts of heterogeneous data across multiple platforms, complicating management. Diverse data formats and distributed sources created significant challenges in aggregation, analysis, and retrieval.

[5] https://www.crewai.com/
[6] https://langchain-ai.github.io/langgraph
[7] https://github.com/Significant-Gravitas/AutoGPT

MontyCloud had multiple APIs which supported various cloud operations and functionalities. The integrated APIs had varying data formats and sources, requiring extensive standardisation efforts for analysis. As features expanded, data volume from these APIs grew rapidly, requiring a system capable of real-time processing of diverse datasets to ensure data completeness and relevance. We faced challenges such as incomplete context retention, data processing inaccuracies, and hallucinations, resulting in irrelevant or incomplete responses. These issues significantly impact complex cloud operations, which depend on precise qualitative insights.

*B. CH2: Maintainability*

As the system expanded, maintaining its functionality became increasingly difficult. The addition of new components and categories made task management and classification challenging, as the delineation of responsibilities among tasks was often unclear. Mapping intent of the query to the functionality became complex as the number of required classification samples grew substantially.

Additionally, as the system evolved, implementing changes or updates became challenging and time-consuming due to the lack of modularity. Integration with external tools required extensive testing to avoid conflicts, thereby slowing down updates and maintenance. Furthermore, Debugging and tracing operational issues became difficult, reducing overall efficiency.

*C. CH3: Extensibility and Modularity*

The monolithic system's dependence on a single vendor's tools and models restricted its flexibility, making it difficult to select the optimal solutions for specific tasks, resulting in a vendor lock-in. The limitation prevented the integration of specialised models tailored to unique tasks and domains, a core principle of Domain-Driven Design (DDD) [10]. This forced developers into a one-size-fits-all approach, constraining the available solutions for diverse CloudOps scenarios.

Moreover, the system's design limited contributions from external developers, as domain experts are not directly involved in the primary workflows. Expanding the system to accommodate other cloud platforms such as Azure or GCP, introduced additional challenges due to the lack of standardised APIs and inconsistent data formats. This vendor-specific monolithic approach complicated API integration and made it difficult to adapt to evolving data structures, hindering the system's maintainability and extensibility.

To address these limitations, we developed an architecture for a multi-agent framework to support the autonomous CloudOps better. Adopting a modular, multi-agent architecture becomes a more effective approach. However, multi-agent systems also presented their own set of engineering challenges such as agent coordination, orchestration, data consistency, context sharing, and importantly, balancing autonomy and control. The following section details how we developed this framework and how it addresses the challenges outlined above.

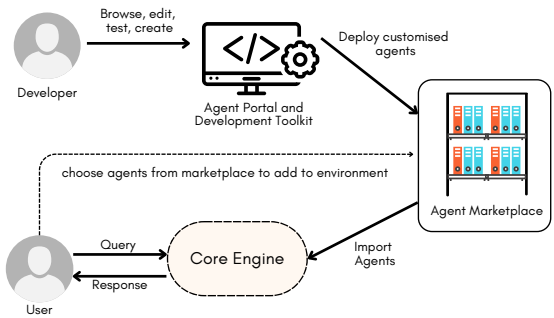

Fig. 3. High-Level Overview of MOYA Framework

## IV. MULTIAGENT FRAMEWORK

MOYA Framework is engineered to mitigate the challenges presented in Section III by creating a collection of modular, interoperable agents that work collaboratively to ensure quality and scalability. Each agent is an autonomous component designed to handle specific tasks in CloudOps, enabling seamless integration with diverse APIs, dynamic data sources, and user-defined workflows, while maintaining runtime flexibility and support for external customisations. Agents can be deployed to monitor operational activities, generate status summaries, and perform various specialised functions within the system.

As illustrated in Figure 3, the MOYA Framework supports the dynamic creation, customisation, deployment, and utilisation of agents, providing a customised approach to CloudOps. It further offers integration with the following set of subsystems:

**Development Toolkit:** Provides developers with tools and resources for building, testing, and integrating agents.

**Agent Portal:** A web-based interface for developers and customers to browse, create, edit, test, and debug agents in a sandbox environment. It simulates the core system's workflows and monitoring invocation probabilities, logs, and metrics.

**Agent Marketplace:** An agent store where developers and users can add or swap in new agents to support specific needs within their cloud environment. Agents can be added by selecting from existing configurations, or, for highly customised needs, by defining additional fields of the agents like Agent URL, description of the agents, functionality, and the permissions the agent would have.

This simplifies the integration of specialised *internal* or *third-party agents*, enabling users to expand their system's capabilities as their needs evolve.

As described in Figure 3, developers can create, test, and deploy agents to the *Agent Marketplace*, simplifying the development process by allowing them to focus solely on functionality. The framework handles underlying complexities like LLM repositories and data streaming, ensuring adaptability. This enables businesses to integrate agents seamlessly into their operations with minimal system changes

In order to manage the MOYA Framework and address the challenges outlined in Section III, we built the *Core Engine* of MOYA as illustrated in Figure 4. As shown in Figure 3, the *Core Engine* is the central component of the MOYA Framework, responsible for driving operations and managing the workflows triggered by user queries or requests. The next subsection provides a detailed view of its architecture.

### A. Core Engine

The Core Engine serves as the central component of the MOYA Framework, handling the query processing, agent coordination, and response generation, as depicted in Figure 3. Figure 4 illustrates its key components and interactions. Additionally, the end-to-end flow of user queries through the Core Engine is visualised in Figure 5.

The engine is triggered by requests or user queries, initiating a coordinated workflow among its key components which are detailed below:

**Request Handler:** The workflow initiates at the *Request Handler*, which processes the query by capturing, validating, and parsing incoming user queries. The validated query is then forwarded to the *Orchestrator* for downstream processing.

**LLM Processor:** The *LLM Processor* interprets user queries and refines responses. As depicted in Figure 5, the *LLM Processor* is utilised twice in a typical workflow: first, to preprocess the initial query, and later, after the *Orchestrator* collects responses from agents, to refine and format the final output as per any specified requirements from the user query.

**Orchestrator:** Handles the internal flow of information within the engine's components and between agents, ensuring data is efficiently and accurately routed. It directs parsed queries received from the *Request Handler* to the relevant components and agents, and also supports predefined workflows to streamline common processing patterns.

**Context Memory:** Stores frequently used responses for quick retrieval, and maintains the contextual information about agents such as current agent statuses. It facilitates efficient information sharing and coordination among agents by temporarily holding retrieved data, responses, and intermediate results. It optimises retrieval by enabling the core engine to respond to requests more efficiently.

**Agent Selector:** The *Orchestrator* directs the user query to the *Agent Selector* to select the right agent for the task. Using Machine learning (ML) techniques, the *Agent Selector* classifies and categorises user queries based on predefined task boundaries, then selects the most relevant agent based on its capabilities and the specific requirements of the query. By clearly delineating each agent's responsibilities within the multi-agent framework, the performance of the classifier has been significantly improved, resulting in more accurate decision-making, thus addressing the challenge discussed in III-B.

**Agent Registry:** Stores crucial agent metadata, including invocation protocols, authentication tokens, access endpoints, functionality details, permissions, and parameters. It ensures agents are properly registered, updated, or removed as needed. When the developer creates an agent, the *registry* is updated with the agent's endpoint, system prompts, and relevant parameters, facilitating future accessibility. The *Orchestrator* relies on the *Agent Registry* to fetch the most recent version of an agent's information. Given the dynamic nature of agent metadata, the *registry* regularly updates and manages configurations to reflect any changes. This ability to adapt to runtime updates enhances the extensibility and modularity of the MOYA Frameworkthereby overcoming III-C.

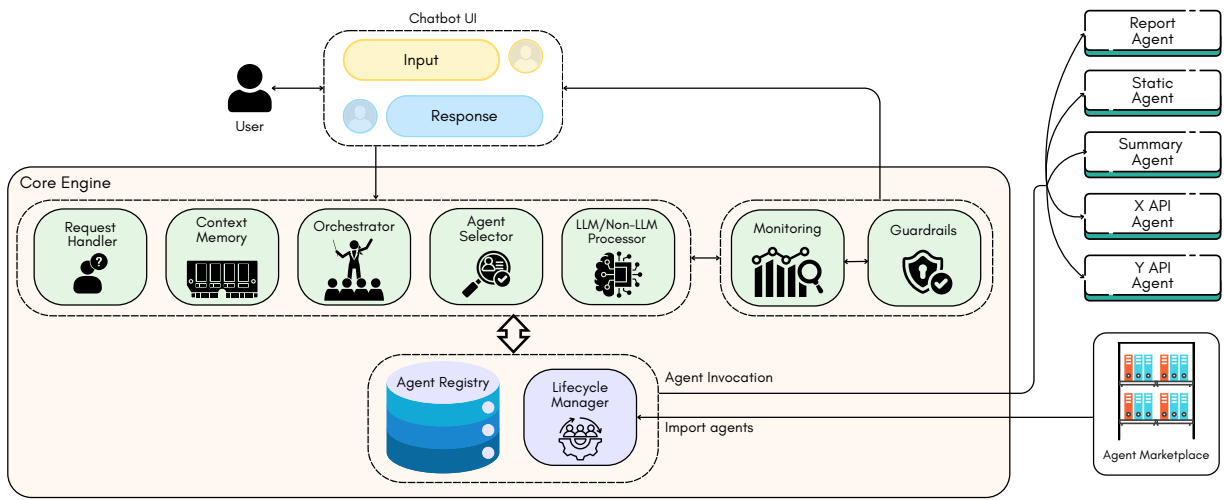

Fig. 4. Detailed View of the Core Engine

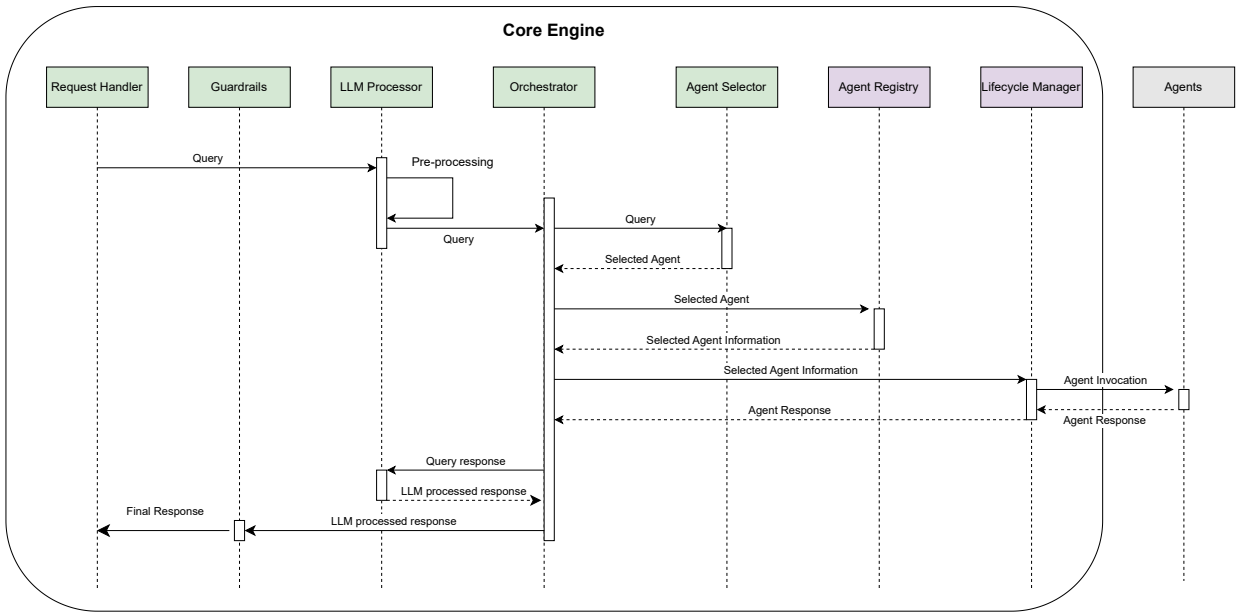

Fig. 5. Interaction flow between Core Engine Components and Agents

**Lifecycle Manager:** The fetched Agent Information is provided as a key to the *Lifecycle Manager* to invoke the respective agent, fetch the response, terminate the agent instance, and forward the response to the *Orchestrator*. It Oversees the life cycle of agents, including invocation, monitoring, and termination. It can also create or import new agents from the *Agent Marketplace*.

**Guardrails and Monitoring:** To mitigate any harmful language or irrelevant information, we employ the concept of AI guardrails [11]. These guardrails are implemented through ML models, as well as deterministic checks that screen responses for potential risks to validity and appropriateness. When a response is flagged by the guardrails system as invalid or potentially harmful, a default response is provided in its place.

To ensure performance, reliability, and operational efficiency we perform offline and runtime monitoring. Offline monitoring incorporates using automated metrics like BLEU and BERTScore as explained in Section VI-B. This ensures the reliability of the system during model or data updates, retraining and new agent integration. Runtime monitoring tracks real-time performance by measuring token usage, response times, and agent invocation patterns and warns users in case of an anomaly. This ensures that MOYA Frameworkmaintains high standards of reliability and safety in its interactions.

## V. AGENTS IN MOYA FRAMEWORK

An agent, as mentioned in Section IV, is dedicated to handle specific tasks in CloudOps. MOYA Framework incorporates both LLM-based and non-LLM-based agents, each tailored to their specific functions. An LLM-based agent tasked with fetching workload summaries can provide targeted insights by interpreting logs and alerting for anomalies, while a non-LLM agent, such as a *Utility Agent*, might focus on generating visualisations or graphs using Python libraries to support reporting tasks.

The modular design of the agents allows for independent development and deployment, minimising disruptions to the overall system and enabling agent reuse across tasks and domains. Drawing inspiration from Domain Driven Design principles [10], we first identified various CloudOps domains we were tackling with, and grouped them into broad categories like APIs, reports, information retrieval, etc. These categories formed the basis for identifying different agents, and establishing clear boundaries around each agent's responsibilities. We were able to define a set number of agent classes distinctly, each accessing only its own data and concerns, leading to stronger cohesion and reduced coupling.

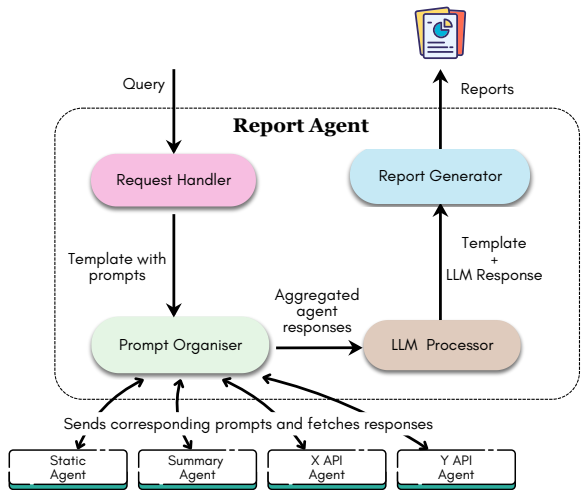

Fig. 6. Report Agent

### A. *Report Agent*

The Report Agent manages the end-to-end process of generating detailed reports based on user queries. Users can trigger report generation through the UI by selecting from available templates or providing a custom template with prompts. For example, a user can request an *"Overview of cloud resources concerning security"*, and the agent generates a comprehensive report with real-time insights from the Cloud system.

When a report is requested, the Report Agent activates the workflow through the *Request Handler*, as illustrated in Figure 6, which captures the query and forwards the template with prompts to the *Prompt Organiser*. The *Prompt Organiser* categorises the prompts and sends them to the relevant agents to retrieve the necessary data. Once the data is fetched, it is passed to the *LLM Processor*, which refines the response to ensure coherence and readability. Finally, the *Report Generator* takes the refined data, integrates it into the initial template, and generates the final report.

Generated reports undergo evaluations through the Core Engine's guardrails to verify correctness and adherence to defined limits. Additionally, the *Report Generator* can create relevant graphs and visual representations, enhancing the readability and utility of the final report. The Report Agent communicates and delegates tasks to other agents, showcasing MOYA's capacity for inter-agent communication and collaboration.

### B. *Static Agent*

The static agent in our system handles queries that require referencing static content which includes all of MontyCloud's documentation and operational guidelines. As depicted in Figure 7 utilizing a RAG approach, this agent processes documents by chunking them, generating embeddings with an embedding model, and storing these embeddings in a vector database. When a query is received, the agent retrieves the most relevant documents with the retrieval engine, and generates contextually appropriate responses. For example, if a user queries about 'Operational Excellence' in AWS's Well-Architected framework, the Agent would retrieve the relevant AWS documentation and generate a response.

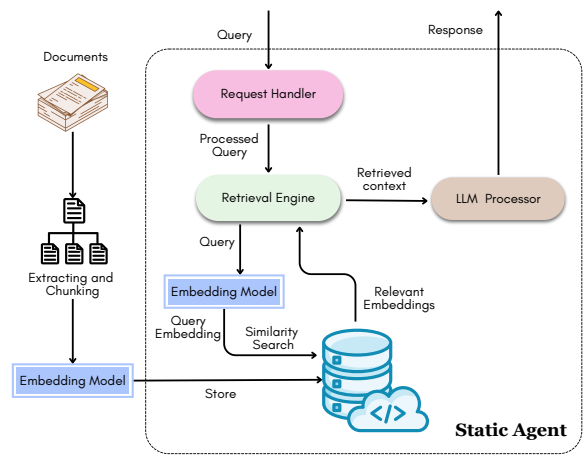

Fig. 7. Static Agent

## C. API Agent: Notification Agent

API agents in MOYA Framework facilitate interaction with both internal and external APIs to retrieve and process relevant information. The generic design of API agents is depicted in Figure 8. The example scenario in Figure 1 demonstrates the integration of external APIs, such as project-management software.

The Notification Agent exemplifies an internal API agent in MOYA Framework. It leverages MontyCloud's APIs to obtain a user's list of unread notifications. It processes and summarises these notifications into a clear and comprehensible format, ensuring users remain efficiently informed about the latest updates and information.

For example, if a user queries about the latest happenings in their account, the Notification Agent would access Monty-Cloud's internal notification-related APIs to retrieve the user's unread notifications. It would then process these notifications into a concise and easily understandable format and give the response back to the user.

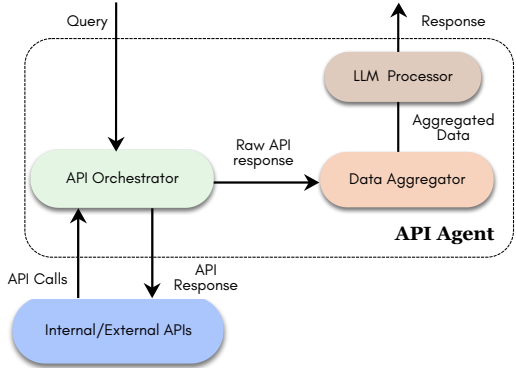

Fig. 8. A general API Agent

## D. Summary Agent

The Summary Agent addresses the requirement for effectively summarising large, distributed cloud workloads(III-A). These workloads consist of data from multiple cloud services and applications, often in varying formats. The data contains critical insights, which are essential for real-time decision-making and efficient cloud management. However, these extensive datasets cannot be processed in a single LLM call due to context length limitations of LLMs, restricting the creation of comprehensive summaries.

Hence, a novel hierarchical summarisation strategy was developed to divide cloud workloads into manageable chunks, each fitting within the LLM's context limits. Each chunk is summarised independently, then compiled together and summarised again into a high-level "summary of summaries". A workload summary like this is generated corresponding to each workload, which is stored and updated at regular intervals.

For instance, when a user requests *"A summary of the changes in my workload over the past month,"* the Summary Agent generates individual summaries for aspects like cost, security, and performance, which are then summarised again to provide a comprehensive overview.

As depicted in Figure 9, when a query about a workload is received, the *Workload Identifier* identifies the workload and sends it to the *Retrieval Engine*, which fetches the relevant summary from the *Summary Store*. The *Pre-processor* combines the summary with the query, and forwards it to the *LLM Processor*, which provides a comprehensive response.

Furthermore, the hierarchical summarisation process was optimised by parallelising the process using multi-threading, enabling concurrent LLM calls. This enhancement allows for a higher frequency of updating the workload summaries, providing the Summary Agent with timely and accurate summaries of large-scale cloud data.

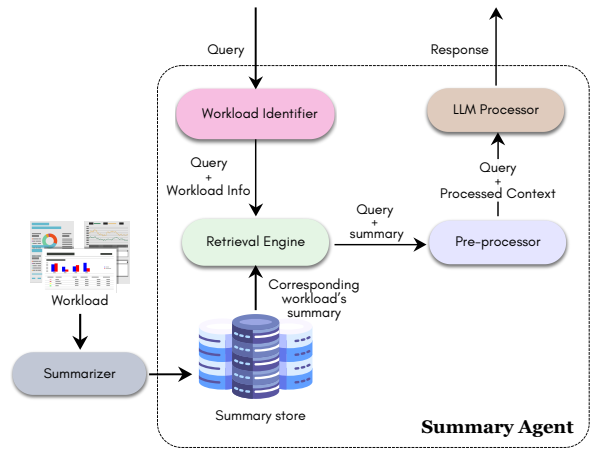

Fig. 9. Summary Agent

## VI. EVALUATION FRAMEWORK

The primary aim of this evaluation is to rigorously assess the performance of our system that leverages the MOYA Framework in real-world scenarios and compare it against our initial RAG-based monolithic implementation. Specifically, this study evaluates the accuracy, relevance, and presentation of responses generated by MOYA Framework, alongside the efficiency in terms of response time and cost.

### A. Experimental Setup

To conduct a comprehensive evaluation, we deployed MOYA Framework in a dedicated development environment to simulate real-world conditions accurately. The system consists of the Core Engine that coordinates the overall framework, and six distinct agents: the Static Agent, Summary Agent, Report Agent, and three API agents—Notification, Cost, and Dashboard. Each agent was implemented as a serverless function, enabling efficient scaling, resource optimisation, and streamlined communication between agents.

For this evaluation, a dataset of 26 prompts was carefully curated to represent a comprehensive range of scenarios relevant to CloudOps. To further assess the system's generalisability and prevent overfitting to specific prompts, 10 variations were generated for each prompt using LLMs, resulting in a total of 260 prompts. These prompts were developed in collaboration with a team of four Product Managers and CloudOps experts from MontyCloud, ensuring their high relevance to practical CloudOps tasks. The prompts encompass multiple domains, including cost analysis, dashboard evaluation, notification management, and general CloudOps queries, thus providing a well-rounded basis for assessment.

In addition, the same expert team curated a high-quality gold standard response for each prompt in the dataset. These expert-verified responses served as the optimal output that the framework should ideally achieve, providing a benchmark for evaluating the framework's performance. By comparing MOYA Framework's outputs against these gold standard responses, we were able to assess its accuracy, relevance, and alignment with CloudOps requirements.

### B. Evaluation Metrics

To thoroughly evaluate the quality of responses produced by MOYA Framework, we employ a combination of widely recognised Natural Language Processing (NLP) metrics and human evaluation, providing a well-rounded assessment of overall performance. For efficiency analysis, we focus on measuring response latency and the cost associated with each interaction by measuring tokens.

*1) Automated Evaluation:* Evaluation of text generation often relies on a combination of metrics rather than a single standard measure. All of these metrics compare how close the generated text is to the golden standard responses. In this line, we use four popular NLP metrics. **ROUGE** (Recall-Oriented Understudy for Gisting Evaluation) [12] is a set of metrics used to evaluate the quality of machine-generated summaries. Here we use rouge-1 which measures the overlap of unigrams between the reference and prediction. The **BLEU** (Bilingual Evaluation Understudy) [13] score is a metric used to evaluate the quality of machine-translated text. **METEOR** (Metric for Evaluation of Translation with Explicit Ordering) [14] is a metric used for evaluating machine-generated text, particularly in the context of machine translation. **BERTScore** [15] is an automatic evaluation metric used to assess the quality of text generation. It leverages pre-trained contextual embeddings from BERT (Bidirectional Encoder Representations from Transformers) [16] and measures the similarity between words in candidate and reference sentences using cosine similarity.

*2) Human Evaluation:* To assess the generalisability of MOYA Framework, a user-centred bug identification session was conducted within the dedicated development environment. This session involved eight team members from MontyCloud, comprising four Product Managers and four Platform Engineers. Each participant used prompts at their discretion, allowing for a diverse evaluation without introducing bias from predefined datasets. Any issues identified during this session were documented as Jira tickets, detailing the system's response, the expected output, and any additional pertinent observations.

To provide a quantitative view of the framework's performance, we categorised the flagged issues into three main categories:

- **Misclassification:** Cases where the monolithic system incorrectly matched prompts to the wrong data source or where MOYA Frameworkdirected the prompt to an incorrect agent.
- **Hallucination:** Instances in which the system generated erroneous information or failed to interpret the prompt accurately.
- **Formatting Issues:** Issues related to incorrect formatting, including errors in tables, bullet points, or other presentation elements.

The distribution of issues within each category provides insight into the framework's performance on a broader scale, highlighting areas of strength and potential improvement in handling diverse user-generated prompts.

*3) Efficiency Analysis:* To assess the efficiency of MOYA Framework, we focus on two primary dimensions: response time and cost (based on input and output token count). These aspects are captured using the following key metrics:

**Total Response Time:** We measure the total response time for both monolithic and multi-agent implementations to compare their performance in delivering timely responses, aiming for minimal response time.

**Token Count:** As token processing incurs costs, we track both input and output tokens to estimate operational costs and evaluate the cost efficiency of each framework configuration.

### C. Results

*1) Automated Evaluation:* The results of the automated evaluation provide insights into the quantitative performance of MOYA Frameworkagainst the baseline monolithic system. Table I summarises the performance across the four selected metrics measured on the dataset of 260 prompts. These metrics offer a multifaceted view of the framework's effectiveness in producing responses closely aligned with the gold standard.

TABLE I
RESULTS

| Approach | rouge-1 | bleu | Meteor | BERTScore | | |
| --- | --- | --- | --- | --- | --- | --- |
| | | | | precision | recall | f1 |
| **Monolith** | 0.321 | 0.102 | 0.265 | 0.854 | 0.834 | 0.843 |
| **MOYA Framework** | **0.448** | **0.221** | **0.423** | **0.867** | **0.869** | **0.868** |

MOYA Frameworkoutperformed the monolithic baseline in key areas, especially in metrics focused on semantic alignment (e.g., METEOR and BERTScore), indicating enhanced interpretative capabilities. A modest improvement in BLEU and ROUGE scores also suggests greater lexical alignment with gold-standard responses. These results suggest that the distributed agent approach supports more accurate responses, achieving closer adherence to target responses.

*2) Human Evaluation:* A total of 22 issues were identified in the monolithic system, compared to 15 issues in MOYA Framework, indicating an overall improvement in performance with the multi-agent framework. For a more granular assessment, each issue was analysed according to its assigned category.

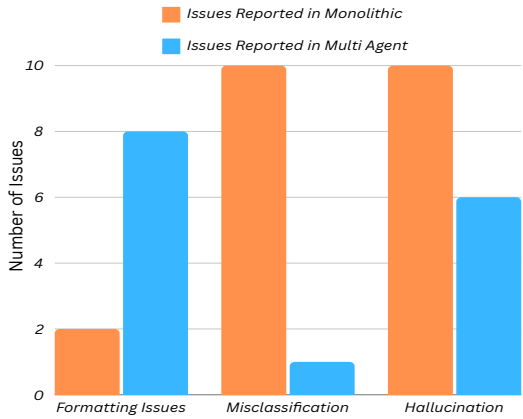

Fig. 10. Issues Reported in Human Evaluation

Figure 10 visualises the distribution of issues flagged by team members during the bug identification session. The issues identified in human testing underscore key areas for improvement, such as response relevance and data inaccuracies.

The transition from a monolithic to a multi-agent framework resulted in a marked reduction in misclassification errors, primarily due to the targeted routing capabilities within individual agents, which improved the system's robustness in handling diverse prompts. However, formatting issues remained the predominant category for MOYA Framework. This challenge was subsequently mitigated by refining system prompts within each agent, resulting in improved response formatting. These findings are instrumental for targeted agent-level enhancements to better support end-user needs and further streamline the framework's output.

*3) Efficiency Analysis:* The multi-agent system, initially expected to be less efficient than the monolithic baseline due to additional LLM calls and API communication overhead, showed surprising performance. Despite higher token consumption for inputs and outputs, the multi-agent framework achieved faster response times, indicating that its decentralised system design may facilitate better code, improving overall performance.

The increased token usage is a tradeoff inherent in the system's design, as multiple agents handle different aspects of the prompt. However, this tradeoff is considered acceptable given the improved response time and processing efficiency. Future work could focus on optimising token usage without compromising latency.

This evaluation highlights the strengths of MOYA Framework in generating accurate and relevant responses fast, while also identifying actionable areas for ongoing enhancement.

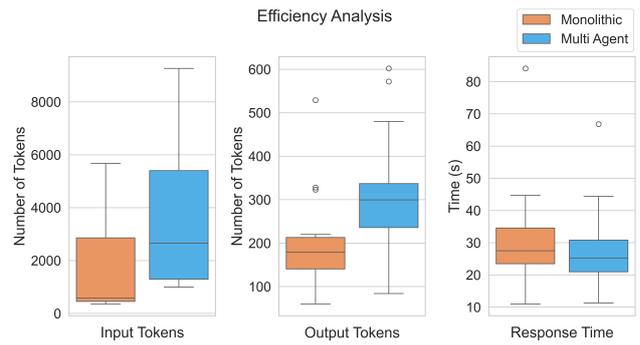

Fig. 11. Efficiency Analysis Results

## VII. DISCUSSION

Table II provides a concise view of how the different challenges were addressed by our XYZ framework. It further highlights the potential of multi-agent GenAI systems in automating CloudOps, demonstrating how a multi-agent framework can address the complexity and specialised needs that simpler function-calling alone cannot effectively handle. The agentic design enables agents to manage distinct, domain-driven tasks across cloud ecosystems, delivering significant benefits in accuracy, and flexibility. This agent-based approach outperforms monolithic models by isolating tasks into specialised agents, streamlining complex CloudOps processes, and enabling adaptability across platforms. These findings align with OpenAI's Level 3 AI paradigm[8], emphasising the value of multi-agent systems in managing and automating complex workflows.

Using distinct agents instead of iterative LLM calls provides targeted domain management that would be difficult to achieve with a single LLM instance. Each agent is dedicated to a discrete domain, ensuring focused processing and avoiding the risks of context overload that iterative LLM calls might incur. This domain specialisation is essential for high-accuracy responses in CloudOps, where different bounded contexts require unique, domain-specific insights.

The framework is built with modularity and cloud-agnostic design to avoid vendor lock-in. While some agents may be specific to certain cloud providers, the overall architecture remains vendor-neutral, ensuring interoperability with AWS, Azure, and GCP, and enabling flexible system evolution without dependency on a single provider.

However, developing and deploying ML-enabled systems in particular multi-agent systems bring unique architecting and engineering challenges on top of the existing challenges highlighted by Lewis et al. [17]. Key obstacles include managing non-determinism, understanding evolving dependencies, and integrating GenAI pipelines, necessitating comprehensive testing frameworks and refined best practices. Traditional software engineering practices, including version control, continuous integration, deployment and testing must adapt to address both the specialised and domain-sensitive requirements of GenAI-powered and agent-specific frameworks.

[8]https://www.youtube.com/watch?v=A6EQp8PL_iE

| Challenge | Primary Cause | Impact | Solution | Resulting Benefit |
| --- | --- | --- | --- | --- |
| Managing Distributed Data | Continuous generation of extensive, distributed data | Difficulty in real-time analysis and insight retrieval | Separation of data sources and data formats for each agent; Specialisation in a single task | Streamlined real-time data processing, enhanced insights, and faster decision-making |
| Maintainability | Monolithic architecture with tight coupling | High technical debt and maintenance challenges | Multi-agent approach for task separation | Independent updates and reduced maintenance time |
| Extensibility and Modularity | Limited flexibility in adding new features | Hindered innovation and integration of new tech | Dynamic Agent Registry for seamless addition of new agents | Faster feature updates and increased adaptability |

TABLE II
CHALLENGES ADDRESSED BY THE MULTI-AGENT FRAMEWORK

Observations from our evaluation reinforce the value of incorporating human oversight in GenAI systems, as suggested by the Copenhagen Manifesto [18] on human-centred AI. Despite the advancements in GenAI and automation, maintaining a human-in-the-loop remains critical, particularly for quality assurance and the nuanced understanding required in complex or sensitive CloudOps tasks. Human insights can effectively supplement automated processes, ensuring relevance, accuracy, and alignment with evolving operational goals.

## VIII. THREATS TO VALIDITY

A potential threat to internal validity arises from the accuracy of agent invocation. Misclassification or improper task delegation could lead to incorrect task execution, thereby impacting the overall performance of the system. While we have implemented a classification mechanism to improve agent invocation accuracy, the increasing number of agents may introduce additional complexity, reducing the precision of task assignment as the system scales.

In terms of external validity, MOYA Frameworkhas primarily been tested within AWS-based cloud environments. As a result, its generalisability to other cloud platforms, such as Microsoft Azure and other user environments, remains uncertain. Variations in cloud service APIs, workflows, and operational contexts could present unforeseen challenges when adapting the system to different or hybrid environments, limiting its broader applicability.

A threat to construct validity comes from the rapid evolution of LLMs. Frequent updates can cause inconsistencies, as newer models may handle tasks differently. This variability undermines the MOYA Framework's consistency and predictability, making it harder to ensure reliable responses across versions. Better engineering practices are needed to be developed to mitigate this issue.

Another threat to construct validity is due to the potential of LLMs hallucinating while generating counts or numbers, such as cloud instance counts or configurations. Although we have implemented validation mechanisms to cross-check critical data points, these measures may not capture all discrepancies.

## IX. RELATED WORK

Alonso et al. [1] present a comprehensive and systematic approach to operationalising cloud environments, addressing notable gaps identified in existing literature. They define essential concepts and challenges associated with cloud operations, proposing a reference framework for CloudOps.

Generative AI, LLMs, and RAG systems are gaining traction in software engineering. Josu et al. [19] explores using LLMs to automate patching in Infrastructure as Code (IaC) projects, incorporating a RAG system to enhance correctness. Eskandani et al. [20] propose using GenAI tools in Functions as a Service (FaaS) for improved monitoring and debugging. Hou et al. [2] discuss the successful use of LLMs in a range of software engineering tasks, emphasising their increasing importance in the field.

Recently, major cloud providers like AWS and Microsoft Azure are developing AI assistants, such as *Amazon Q* [9] and *Microsoft Copilot* [10], to support cloud operations, including code generation, task planning, and troubleshooting. However, these solutions are platform-specific, whereas our framework is designed to work platform agnostic, highlighting a key difference in architecture.

As described in Section IV, many Multi-Agent frameworks have emerged in recent times like AutoGen [7], CrewAI, LangGraph, and AutoGPT. However, these frameworks are not flexible and don't match our requirements of a versatile CloudOps framework.

## X. CONCLUSION AND FUTURE WORK

In conclusion, the MOYA Frameworkoffers a practical approach to CloudOps by integrating multi-agent systems and LLM-powered automation to manage complex cloud operations. Designed to handle diverse tasks, the framework enables real-time insights, improved maintainability, and enhanced flexibility across cloud environments. RAG which combines LLMs with information retrieval techniques, is a key feature of the system that enhances the framework's ability to provide precise insights based on real-time, heterogeneous data. Evaluation results show that the MOYA Frameworkoutperforms traditional monolithic RAG approaches in effectiveness, and task delegation, reducing misclassification and hallucination.

Future work will focus on extending the capabilities of the system's agent interactions to include more sophisticated inter-agent communication and collaborative task handling. We will focus on enhancing inter-agent collaboration, specifically by developing mechanisms for simultaneous response aggregation from multiple agents. This will allow the framework to handle complex queries that require input from different agents, increasing its capacity for comprehensive, real-time insights. Another area for development involves broadening the framework's applicability beyond AWS-specific environments to support hybrid and multi-cloud ecosystems, enhancing its interoperability with Azure, GCP, and other cloud platforms.

---
[9] https://aws.amazon.com/q/
[10] https://azure.microsoft.com/en-us/products/copilot

## REFERENCES

[1] J. Alonso, L. Orue-Echevarria, and M. Huarte, "Cloudops: Towards the operationalization of the cloud continuum: Concepts, challenges and a reference framework," *Applied Sciences*, vol. 12, no. 9, p. 4347, 2022.

[2] X. Hou, Y. Zhao, Y. Liu, Z. Yang, K. Wang, L. Li, X. Luo, D. Lo, J. Grundy, and H. Wang, "Large language models for software engineering: A systematic literature review," 2024.

[3] P. Lewis, E. Perez, A. Piktus, F. Petroni, V. Karpukhin, N. Goyal, H. Küttler, M. Lewis, W. tau Yih, T. Rocktäschel, S. Riedel, and D. Kiela, "Retrieval-augmented generation for knowledge-intensive nlp tasks," 2021.

[4] S. Han, Q. Zhang, Y. Yao, W. Jin, Z. Xu, and C. He, "Llm multi-agent systems: Challenges and open problems," *arXiv preprint arXiv:2402.03578*, 2024.

[5] Y. Shoham and K. Leyton-Brown, *Multiagent Systems: Algorithmic, Game-Theoretic, and Logical Foundations*. USA: Cambridge University Press, 2008.

[6] Y. Shoham and K. Leyton-Brown, *Multiagent systems: Algorithmic, game-theoretic, and logical foundations*. Cambridge University Press, 2008.

[7] Q. Wu, G. Bansal, J. Zhang, Y. Wu, B. Li, E. E. Zhu, L. Jiang, X. Zhang, S. Zhang, A. Awadallah, R. W. White, D. Burger, and C. Wang, "Autogen: Enabling next-gen llm applications via multi-agent conversation," in *COLM 2024*, August 2024.

[8] S. Martínez-Fernández, J. Bogner, X. Franch, M. Oriol, J. Siebert, A. Trendowicz, A. M. Vollmer, and S. Wagner, "Software engineering for ai-based systems: a survey," *ACM Transactions on Software Engineering and Methodology (TOSEM)*, vol. 31, no. 2, pp. 1–59, 2022.

[9] J. Bosch, I. Crnkovic, and H. H. Olsson, "Engineering ai systems: A research agenda," 2020.

[10] Evans, *Domain-Driven Design: Tacking Complexity In the Heart of Software*. USA: Addison-Wesley Longman Publishing Co., Inc., 2003.

[11] Y. Dong, R. Mu, G. Jin, Y. Qi, J. Hu, X. Zhao, J. Meng, W. Ruan, and X. Huang, "Building guardrails for large language models," 2024.

[12] C.-Y. Lin, "ROUGE: A package for automatic evaluation of summaries," in *Text Summarization Branches Out*, (Barcelona, Spain), pp. 74–81, Association for Computational Linguistics, July 2004.

[13] K. Papineni, S. Roukos, T. Ward, and W.-J. Zhu, "Bleu: a method for automatic evaluation of machine translation," in *Proceedings of the 40th Annual Meeting on Association for Computational Linguistics*, ACL '02, (USA), p. 311–318, Association for Computational Linguistics, 2002.

[14] A. Lavie and A. Agarwal, "Meteor: an automatic metric for mt evaluation with high levels of correlation with human judgments," in *Proceedings of the Second Workshop on Statistical Machine Translation*, StatMT '07, (USA), p. 228–231, Association for Computational Linguistics, 2007.

[15] T. Zhang, V. Kishore, F. Wu, K. Q. Weinberger, and Y. Artzi, "Bertscore: Evaluating text generation with bert," 2020.

[16] J. Devlin, M.-W. Chang, K. Lee, and K. Toutanova, "Bert: Pre-training of deep bidirectional transformers for language understanding," 2019.

[17] G. A. Lewis, H. Muccini, I. Ozkaya, K. Vaidhyanathan, R. Weiss, and L. Zhu, "Software Architecture and Machine Learning (Dagstuhl Seminar 23302)," *Dagstuhl Reports*, vol. 13, no. 7, pp. 166–188, 2024.

[18] D. Russo, S. Baltes, N. van Berkel, P. Avgeriou, F. Calefato, B. Cabrero-Daniel, G. Catolino, J. Cito, N. Ernst, T. Fritz, H. Hata, R. Holmes, M. Izadi, F. Khomh, M. Kjærgaard, G. Liebel, A. Lluch-Lafuente, S. Lambiase, W. Maalej, and B. Vasilescu, "Generative ai in software engineering must be human-centered: The copenhagen manifesto," *Journal of Systems and Software*, vol. 216, p. 112115, 05 2024.

[19] J. Diaz-de Arcaya, J. López-de Armentia, G. Zárate, and A. I. Torre-Bastida, "Towards the self-healing of infrastructure as code projects using constrained llm technologies," in *Proceedings of the 5th ACM/IEEE International Workshop on Automated Program Repair*, pp. 22–25, 2024.

[20] N. Eskandani and G. Salvaneschi, "Towards ai for software systems," in *Proceedings of the 1st ACM International Conference on AI-Powered Software*, pp. 79–84, 2024.